\documentclass[prd,nofootinbib,showpacs,reprint]{revtex4}
\usepackage[a4paper, total={6.5in, 8.5in}]{geometry}
\usepackage{graphicx}
\usepackage{bm}
\usepackage[usenames,dvipsnames]{color}
\usepackage{amsmath,amssymb}
\usepackage{slashed}
\usepackage{float}
\usepackage{epsfig}
\usepackage{subfigure}
\usepackage{latexsym, amssymb} 
\usepackage[colorlinks=true,linkcolor=blue,citecolor=blue]{hyperref}
\usepackage[normalem]{ulem}

\begin{document}
	
	\title{
  Composite Dissipation in Warm Inflation: Implications for the Primordial Power Spectrum
  }
   \author{Ayush Sahu}
	\email{ayush.2024rph02@mnnit.ac.in}
	
	\affiliation{Department of Physics, Motilal Nehru National Institute of Technology Allahabad, Prayagraj 211004, U.P., India}
	 
\author{Richa Arya}
	\email{richaarya@mnnit.ac.in} 	\affiliation{Department of Physics, Motilal Nehru National Institute of Technology Allahabad, Prayagraj 211004, U.P., India}
	\author{Sergio E. Jor\'as}
	\email{joras@if.ufrj.br}
	\affiliation{
		Instituto de F\'\i sica, Universidade Federal do Rio de Janeiro,\\
		CEP 21941-972 Rio de Janeiro, RJ, Brazil}
\author{Karim H. Seleim}
\email{karim.seleim@uofcanada.edu.eg}
\affiliation{School of Mathematical and Computational Sciences, Universities of Canada in Egypt (UCE), New Administrative Capital, Cairo, Egypt}
\affiliation{
University of Science and Technology, Zewail City of Science and Technology, 6th of October City, Giza 12588, Egypt
}
	
	\date{\today}

	\begin{abstract}
    Warm inflation is a well-motivated and generalized framework of inflation, 
   describing a 
   coupled inflaton-radiation bath.
	In this work, we investigate a warm inflation model with  a quartic potential and a composite dissipation coefficient 
	$\Upsilon(\phi, T) = C_1 \frac{T^3}{M_{\text{Pl}}^2} + C_2 \frac{T^3}{\phi^2}.$
	The two terms in $\Upsilon$ dominate at different scales: the first term governs the early inflationary dynamics at large (CMB) scales, while the second term becomes significant at smaller scales. The model features two distinct stages of inflation: an initial phase where strong dissipation ($Q \gg 1$) generates a red-tilted primordial spectrum consistent with CMB observations (from ACT), followed by a second phase producing a blue-tilted spectrum with a significant amplification of power at small scales, leading to primordial black hole formation.
	We analyze the effects of key parameters -- like the duration of each inflationary phase, the slow-roll parameter at the end of the first phase, the dissipation strength at the pivot scale, and the choice of the growth function -- on the primordial power spectrum and its spectral index. 
    Additionally, we examine the consistency of the model with the swampland distance conjecture and trans-Planckian conjecture, needed for embedding these models with some UV complete theories. This work highlights the potential of warm inflation with a composite dissipation coefficient to reconcile large-scale CMB measurements with small-scale structure formation.
		
	\end{abstract}

	\maketitle
	
	\section{Introduction}
    The inflationary paradigm \cite{Kazanas:1980tx,Sato:1980yn,Guth:1980zm,Linde:1981mu} of the early Universe efficiently solves the problems faced by the Standard Model of cosmology, as well as explains the CMB observations very precisely. Additionally, it provides a mechanism for generating the primordial density inhomogeneities that later evolve into the seeds of the Large-Scale Structure (LSS) of the Universe, further reinforcing its significance and success.

In the standard cold inflation description, it is presumed that the inflaton’s coupling
to other fields are ineffective during the inflationary phase. Therefore, because of the
nearly exponential expansion of the Universe during inflation, the number densities
of all the species present at that epoch dilute away, and the Universe enters into a
supercooled state. Further, when the inflationary phase ends, the Universe undergoes
a (p)reheating phase in which the inflaton oscillates and decays into particles. For reviews on  cold inflation, see Refs. \cite{Baumann:2009ds,Linde:2007fr,Olive:1989nu,Riotto:2002yw,Tsujikawa:2003jp,Linde:2005ht}.

Warm inflation \cite{Berera:1995wh,Berera:1995ie,Berera:1998px} is a generalized framework, in which the inflaton field continuously couples and dissipates its energy into a thermal bath of radiation during the accelerated expansion. As a result, the universe remains at a finite temperature throughout, eliminating the need for a distinct reheating phase to initiate the radiation-dominated era. The state of the universe in warm inflation is not the quantum vacuum, as in the standard cold inflation, but rather a statistical thermal bath. Due to the presence of temperature,  both the background dynamics and the generation of primordial fluctuations are modified.  For reviews on warm inflation, see Refs. \cite{Berera:2006xq,Berera:2008ar,Campo11,Oyvind&nbspGron:2016zhz,Kamali:2023lzq}.

There are many attractive features in warm inflation: firstly, it is more natural to consider inflaton interactions with other fields, even if small. In the limiting case when the interactions are neglected, one can always recover the standard cold description.  As the universe has a temperature throughout the warm inflationary phase, a separate reheating phase might not be needed at the end. Thus, warm inflation
evades the \textit{graceful exit} problem faced by cold inflation \cite{Das:2020lut,Kamali:2023lzq}. Studies show that the tensor-to-scalar ratio in certain warm inflationary models is reduced, therefore,
 some cold inflation models ruled out from the CMB observations are allowed within the warm description \cite{Visinelli:2016rhn,Benetti:2016jhf,Arya:2017zlb,Bastero-Gil:2017wwl,Arya:2018sgw}. As the presence of dissipation modifies the slow-roll conditions in warm inflation, thus, steeper potentials of inflation can also be allowed in this framework. Hence, the \textit{$\eta$-problem} can be relaxed with a strong dissipation in warm inflation models \cite{Berera:2004vm}. Further, warm inflationary models also lead to unique and significant non-Gaussianities, which if observed in future, would provide a test to these models \cite{Gupta:2002kn,Moss:2011qc,Bastero-Gil:2014raa,Mirbabayi:2022cbt}. As inflation is an effective theory, it has to be embedded in some UV theory, which is ensured through some swampland and trans-Planckian conjectures. In some studies, it is shown that the warm inflationary models in strong dissipative regime follow these conjectures \cite{Das:2018rpg,Motaharfar:2018zyb}, thereby implying that they lie in the landscape of UV complete theories.

 Another  crucial and recently explored aspect of warm inflationary models is the presence of features at the small-scales. In Ref. \cite{Arya:2019wck}, it was showed for the first time that warm inflation can lead to an enhanced curvature power spectrum at the small-scales, which is a result of a growing dissipation parameter during inflation.  This enhancement occurs by several orders of magnitude and leads to the formation of primordial black holes (PBH). In contrast to existing studies in cold inflation where the desired enhancement is generated by imposing a feature in the inflationary potential (like a bump/dip, or an inflection point) \cite{Germani:2017bcs,Bhaumik:2019tvl,Mishra:2019pzq,Ragavendra:2023ret}, our study shows that warm inflation naturally leads to such an effect. This conclusion has also been supported by further investigations into other warm inflation models, as discussed in Refs. \cite{Bastero-Gil:2021fac,Correa:2022ngq,Correa:2023whf}.
Moreover, it is also pointed out that the PBH can also constitute a fraction of dark matter abundance. In a certain mass window, PBHs may account for the entirety of the dark matter abundance. In our study in Ref. \cite{Arya:2023pod}, we investigated a warm inflation model with a non-linear Galileon-like kinetic term and  and demonstrated that such models can indeed generate PBHs consistent with the total dark matter composition. Additionally, large scalar fluctuations also couple to the tensor fluctuations at the second order, and lead to a spectrum of scalar induced gravitational waves. 
 In Refs. \cite{Bastero-Gil:2021fac,Arya:2022xzc}, it is shown that a spectrum of ultrahigh-frequency induced gravitational waves is obtained from warm inflation models.

The objective of this work is to investigate the effects of a composite dissipation coefficient—comprising two distinct terms that dominate at different energy scales—on the dynamics and observational signatures of warm inflation. 
The dissipation coefficient arises from the interaction of the inflaton with an ambient environment of fields. The microphysical origin of dissipation depends on several factors, including the decay channels of the inflaton into radiation, the coupling strength with other fields, the masses and multiplicities of the radiation fields, and the temperature of the thermal bath. As the inflationary dynamics evolve, different forms of the dissipation coefficient may become dominant at different stages, leading to a scale-dependent dissipation mechanism that can imprint distinct signatures on the primordial power spectrum.
We note that a scale-dependent dissipation coefficient has also been discussed in Refs.~\cite{BasteroGil:2010pb,Bastero-Gil:2021fac}, where different forms are approximated in the high-temperature and low-temperature limits, respectively. Our study differs significantly from Ref.~\cite{Bastero-Gil:2021fac}, where inflation begins in the weak dissipation regime with $\Upsilon \propto 1/T$, and transitions to the strong dissipation regime characterized by $\Upsilon \propto T^k$.  Also, in Ref. ~\cite{BasteroGil:2010pb}, at low temperatures, the leading contribution to the dissipation coefficient is $\Upsilon \propto T^3/M^2$ while at high temperatures, it is $\Upsilon \propto T$.  In contrast, our model 
has initially $\Upsilon \propto T^3/M_{\text{Pl}}^2$ dominantly, which naturally supports strong dissipation during the early (CMB scale) phase of inflation, and then takes the form $\Upsilon \propto T^3/\phi^2$ that continues to maintain strong dissipation while enhancing power at the small-scales. 
As a consequence of sustained strong dissipation, this model features a sub-Planckian field excursion and remains consistent with the swampland distance conjecture. Moreover, the tensor-to-scalar ratio is highly suppressed in our model, distinguishing our results sharply from these studies.

This paper is organized as follows: In Section~\ref{WI}, we provide a brief overview of the fundamentals of warm inflation, including the slow-roll parameters and conditions, the dissipation coefficient, and the primordial power spectrum. In Section~\ref{composite}, we introduce our warm inflation model featuring a composite dissipation coefficient, and outline the corresponding dynamical equations for each phase of inflation, along with the numerical framework used in this study. We then present our results in Section~\ref{results}, focusing on the impact of various model parameters on CMB observables. In Section~\ref{back}, we show the background evolution of our composite dissipation warm inflation models and discuss the swampland conjectures. Finally, we summarize our findings and conclusions in Section~\ref{summary}.

    \section{Brief Overview of Warm Inflation}
	\label{WI}
	The framework of warm inflation considers energy dissipation during inflation, using the tools of non-equilibrium field theory for interacting quantum systems. Within this setup, it is typically assumed that the inflaton evolves slowly and stays close to thermal equilibrium, as required by the adiabatic approximation \cite{Gleiser:1993ea,Berera:1998gx,Berera:2007qm}. 
The background evolution of the inflaton field $\phi$ along its potential $V(\phi)$ during warm inflation is altered by a dissipation term $\Upsilon\dot{\phi}$, introduced due to interactions between the inflaton and other fields. The equation of motion then becomes:
	\begin{equation}
		\ddot{\phi}+3H\dot{\phi}+\Upsilon\dot{\phi}=-V_{,\phi}.
		\label{phieom}
	\end{equation}
	Here, a dot denotes differentiation with respect to cosmic time $t$, $H$ is the Hubble parameter, and the subscript $_{,\phi}$ indicates a derivative with respect to $\phi$. The dissipation coefficient $\Upsilon(\phi, T)$ generally depends on the inflaton field value and the cosmic temperature. Using the dimensionless dissipation parameter $Q \equiv \frac{\Upsilon}{3H}$, Eq.~(\ref{phieom}) can be rewritten as:
	\begin{equation}	
		\ddot \phi + 3 H( 1+ Q ) \dot\phi + V_{,\phi}=0.
		\label{inflatoneom}
	\end{equation}
	The dissipation parameter $Q$ is a dimensionless quantity, with $Q > 1$ characterizing the strong dissipative regime and $Q < 1$ defining the weak dissipative regime of warm inflation.

	The dissipation of the inflaton results in a transfer of energy to the radiation component, which is expressed as:
	\begin{equation}
		\dot\rho_r+4H\rho_r=\Upsilon{\dot\phi}^2~.
		\label{rad}
	\end{equation}
	It is considered that the radiation produced through dissipation thermalizes rapidly, thus, $\rho_r = \frac{\pi^2}{30} g_* T^4$, where $T$ denotes the temperature of the thermal bath and $g_*$ is the effective number of relativistic degrees of freedom during warm inflation. It is important to note that while the energy-momentum tensors of the inflaton and radiation components are not individually conserved due to energy exchange, the total energy-momentum tensor of the combined system remains conserved.
	\subsection{Slow-roll parameters and conditions}
	To ensure a sufficient period of inflation, the inflaton potential must be nearly flat, allowing the field to evolve slowly. This can be characterised by small slow-roll parameters:
	\begin{equation}
		\epsilon_\phi = 
		\frac{M_{\text{Pl}}^2}{16 \pi}\,\left(\frac{V_{,\phi}}{V}\right)^2, \hspace{1cm}
		\eta_\phi = 
		\frac{{M_{\text{Pl}}^2}}{8\pi}
		\,\left(\frac{V_{,\phi\phi}}{V}\right)
		\label{coldslow}
	\end{equation}
	where $M_{\text{Pl}}\equiv\sqrt{\frac{1}{G_N}} \simeq 1.22\times10^{19}$ GeV is the  Planck mass and $G_N$ is the gravitational constant.
	In warm inflation, there are additional slow-roll parameters, which measure the field and temperature dependence of the inflaton potential and the dissipation coefficient \cite{Hall:2003zp,Moss:2008yb}
	\begin{equation}
		\beta_\Upsilon = 
		\frac{{M_{\text{Pl}}^2}}{8\pi}
		\,\left(\frac{\Upsilon_{,\phi}\,V_{,\phi}}{\Upsilon\,V}\right),  \hspace{0.5cm}
		b=\frac{T V_{,\phi T}}{V_{,\phi}}~,\hspace{0.5cm} c=\frac{T\Upsilon_{,T}}{\Upsilon}.
		\label{warmslow}
	\end{equation}
	Here the subscript $_{,T}$ represents derivative  with respect to $T$. The stability analysis of warm inflation leads to the following conditions on these slow-roll parameters \cite{Moss:2008yb}
	\begin{align} 
		&\epsilon_\phi \ll 1+Q,\hspace{0.8cm} |\eta_\phi| \ll 1+Q,\hspace{0.8cm} |\beta_\Upsilon| \ll 1+Q, 
		\hspace{0.8cm} 0<b\ll\frac{Q}{1+Q}, \hspace{0.8cm} |c|\leq 4.
		\label{slow_roll}
	\end{align}
	We can see that the conditions on the slow-roll parameters $\epsilon_\phi$ and $\eta_\phi$ in warm inflation are relaxed, as the upper limits on these parameters are $1+Q$ rather than $1$ in cold inflation. Physically, the slow roll is sustained by a strong dissipation even under a steep potential.
	Consequently, with a large dissipation parameter $Q$, warm inflation description  helps to ease the $\eta$-problem \cite{Berera:2004vm}.
	The condition on $b$ implies that warm inflation is only feasible when the thermal corrections to the inflaton potential remain small.

	In the slow-roll approximation, we can neglect $\ddot \phi$ in Eq. (\ref{inflatoneom}), which leads to
	\begin{equation}
		\dot\phi\approx \frac{-V_{,\phi}}{3H(1+Q)},
		\label{phido}
	\end{equation}
	Additionally, assuming that the radiation energy density evolves slowly, we
	can approximate $\dot\rho_r\approx 0$ in Eq. (\ref{rad}) and obtain 
	\begin{equation}
		\rho_r 
		\approx \frac{\Upsilon}{4H} {\dot\phi}^2 =\frac{3}{4} Q {\dot\phi}^2.
		\label{rhor}
	\end{equation}	
This indicates that the radiation energy density is directly controlled by the dissipation parameter $Q$, as expected in warm inflation. The dissipation of the inflaton field to radiation continues throughout the duration of inflation, 
	which comes to an end when either the energy density of the radiation bath becomes larger than that of the inflaton or the slow-roll conditions are violated.
    
	\subsection{Dissipation coefficient}
	The dissipation coefficient $\Upsilon(\phi, T)$ arises from the underlying microphysics of the coupled inflaton-radiation system. It depends on factors such as the inflaton decay channels, its coupling strength to other fields, the masses and multiplicities of the radiation fields involved, and the temperature of the thermal bath. See, for instance, Refs. \cite{Moss:2006gt,BasteroGil:2010pb,BasteroGil:2012cm,Bastero-Gil:2016qru,Bastero-Gil:2018yen,Bastero-Gil:2019gao,Berghaus:2019whh} for the finite-temperature quantum-field theory calculations of the dissipation coefficient. 
    
     A general form of dissipation coefficient used in the literature is:
	$$
	\Upsilon(\phi, T) = C_\Upsilon T^c \phi^p M^{1-p-c},		$$
	where  $C_\Upsilon$, \(c\), \(p\), and \(M\) are model-dependent constants determined by the specifics of the underlying interactions \cite{Motaharfar:2018zyb,Kamali:2023lzq,Montefalcone:2023pvh,Santos:2024pix}. 
    The most commonly studied forms of dissipation coefficient are  \cite{Benetti:2016jhf,Bastero-Gil:2017wwl,Arya:2017zlb,Arya:2018sgw,Motaharfar:2018zyb,Ballesteros:2023dno}:
    \begin{align}
    &C_\Upsilon \frac{T^3}{\phi^2},\\
    &C_\Upsilon T ,\qquad  {\rm and}\\
    &C_\Upsilon \frac{T^3}{M^2}.
    \end{align} 
    The first expression 
    is motivated from two-stage decay of the inflaton in a supersymmetric inflation model \cite{Moss:2006gt,BasteroGil:2010pb}. It arises in the low-temperature limit, where the temperature of the thermal bath is much less than the masses of the intermediate catalyst fields. 
The second one 
is also obtained in a supersymmetric warm inflation in the high-temperature limit, when the mass of the intermediate field is smaller than the temperature of thermal bath.
This form of dissipation coefficient also arises in the ``Little Higgs" models of electroweak symmetry breaking with Higgs as psuedo-Nambu Goldstone boson of a broken $U(1)$ gauge symmetry \cite{Bastero-Gil:2016qru}. The latter 
is obtained in minimal warm inflation model where the inflaton has an axion-like coupling to
Yang-Mills gauge fields \cite{Berghaus:2019whh}.

\subsection{Primordial power spectrum}
The primordial curvature power spectrum in warm inflation receives contributions from both quantum and thermal fluctuations of the inflaton field, and is given by Refs. \cite{Hall:2003zp, Graham:2009bf, Bastero-Gil:2010dgy}:
\begin{equation}
    P_\mathcal{R}(k) = \left(\frac{H_k^2}{2\pi\dot\phi_k}\right)^2 \left[1 + 2n_k + \left(\frac{T_k}{H_k}\right)
    \frac{2\sqrt{3}\pi Q_k}{\sqrt{3 + 4\pi Q_k}}\right] G(Q_k),
    \label{Pkfull}
\end{equation}
where, the subscript $k$ denotes that all quantities are evaluated at the time when the $k^{\text{th}}$ mode crosses the horizon during inflation.
Here, the term $\left(\frac{H_k^2}{2\pi\dot\phi_k}\right)^2$ corresponds to the standard quantum vacuum contribution as in cold inflation. The additional terms inside the square brackets represent enhancements due to thermal fluctuations that arise in the presence of a thermal bath during warm inflation. 
 In a thermal environment, the inflaton field can be excited above its vacuum state, and $n_k$ accounts for the inflaton particle number density. It is typically modeled by the Bose-Einstein distribution, $n_k = n_{\text{BE}}$.

A key feature of warm inflation is that the temperature of the thermal bath remains larger than the Hubble parameter, $T/H > 1$, throughout the inflationary phase, ensuring that the radiation remains thermalized even in an expanding background. In this context, the thermal noise contribution becomes dominant, especially in the strong dissipative regime ($Q \gg 1$).
Moreover, when the dissipation coefficient has a temperature dependence of the form $\Upsilon \propto T^c$, radiation fluctuations can feed back into the inflaton fluctuations, modifying the growth of perturbations, as described in Refs.~\cite{Bastero-Gil:2016qru, Bastero-Gil:2018uep}. This effect is encapsulated by the function $G(Q)$. Above, it acts as an enhancement (or suppression) factor depending on the sign of $c$ -- for $c > 0$, $G(Q)$ enhances the spectrum, while for $c < 0$, it suppresses it. The different forms of $G(Q)$ obtained in the literature are given below \cite{Bastero-Gil:2016qru,Benetti:2016jhf,Bastero-Gil:2018uep}:
\begin{equation*}
G(Q)=1+ 0.335~Q^{1.364}+ 0.0185~Q^{2.315} \qquad \textrm{for $c=1 ~(\Upsilon\propto T)$}\end{equation*}
\begin{equation*}
G(Q)=1+ 4.981~ Q^{1.946}+ 0.17 ~Q^{4.330} \qquad \textrm{for $c=3 ~(\Upsilon\propto T^3)$}
\end{equation*}
\begin{equation*}
G(Q)=\frac{1+1.30 ~Q^{0.1}}{(1+0.25~ Q^{0.79})^{1.98}} \qquad \qquad \qquad \textrm{for $c=-1 ~(\Upsilon\propto 1/T)$}.
\end{equation*}

	\section{Warm Inflation with Composite dissipation coefficient}
	\label{composite}
    Having established the basic framework of warm inflation, we now introduce
 Composite Dissipation Warm Inflation (CDWI), defined as a combination of two forms of dissipation coefficient
	\begin{equation}
		\Upsilon(\phi,T)
        =         C_1 \frac{T^3}{M_{\text{Pl}}^2}  +
         C_2 \frac{T^3}{\phi^2}.		\label{eq:dissipation}
	\end{equation}
	The relative dominance of each term governs the transition between the two inflationary phases. The first term dominates during the initial phase (Phase-I), which lasts for $N_1=35-40$ e-folds, while the second term becomes significant in the subsequent phase of inflation (Phase-II), which lasts for $N_2=10-15$ e-folds.

Both forms of the dissipation coefficient have been studied individually in the literature~\cite{Berghaus:2019whh,PhysRevD.101.103529,Kamali:2021ugx,Arya:2017zlb,Ballesteros:2023dno}, and their dynamics are well understood.
The first term, $\Upsilon \propto \frac{T^3}{M_{\text{Pl}}^2}$, is known to support a strong dissipation regime consistent with the CMB observations. This allows for a sub-Planckian inflaton field excursion, which is favorable in light of the swampland conjectures. On the other hand, the second form, $\Upsilon \propto T^3/\phi^2$
leads to a rapid evolution of the dissipation parameter, resulting in a blue-tilted primordial spectrum at small scales.
These features indicate the idea of a composite dissipation coefficient in which inflationary dynamics smoothly transitions from one regime to the other. With such a setup, we can construct a self-consistent warm inflation model which allows a strong dissipative regime throughout and a natural enhancement in the primordial power spectrum at the small-scales, leading to a generation of primordial black holes. With this motivation, we try combinations of different forms of dissipation coefficient and see that our CDWI model Eq. (\ref{eq:dissipation}) is a consistent choice.

From a physical standpoint, dissipation arises due to interactions between the inflaton and a bath of radiation or intermediate fields, and the precise functional form of the dissipation coefficient depends on the microscopic details of these interactions. As the inflaton evolves during inflation, its coupling structure and the mass hierarchy of the interacting fields can dynamically vary. This evolution can naturally lead to a shift in the dominant 
dissipation mechanism, 
resulting in a smooth transition between different functional forms of the dissipation coefficient. Such interpolations are not only plausible but are also expected in the framework of effective field theory, where different operators become relevant at different energy scales. Therefore, a composite dissipation coefficient can be interpreted as an effective description of a more complex underlying particle physics model, capturing the scale-dependent structure of inflaton-radiation interactions.

    Now we describe the dynamical equations governing both the phases of warm inflation. We consider CDWI model with a quartic potential 
    \begin{equation}
    V(\phi) = \lambda \phi^4,
\end{equation}
where $\lambda$ is the self-coupling of inflaton field.
    As during inflation, the energy density of inflaton dominates the total energy density of the universe, the Hubble parameter for the universe is given as
    \begin{equation}
        H^2\approx\frac{8\pi}{3 M_{\text{Pl}}^2} V(\phi)=\frac{8\pi}{3 M_{\text{Pl}}^2} \lambda\phi^4.
        \label{Hubble}
    \end{equation}
    In the slow-roll approximation, we get from Eq. (\ref{phido})
\begin{equation}
    \dot\phi\approx-\frac{4}{3} \sqrt{\frac{3}{8\pi}}\sqrt{\lambda}\frac{\phi M_{Pl}}{(1+Q)}
    \label{phidot}.
\end{equation}
Thus, the prefactor term in the power spectrum becomes
\begin{equation}
\frac{H_k^2}{2\pi\dot\phi}=
\sqrt{\frac{8\pi}{3}}\sqrt{\lambda}
\left(\frac{\phi}{M_{\text{Pl}}}\right)^3 (1+Q). 
\label{front}
\end{equation}
Further, for the thermal radiation bath,  energy density $\rho_r=\frac{\pi^2 g_*T^4}{30}$, where $g_*$ is the number of relativistic degrees of freedom during warm inflation, which we consider as
$g_*\approx 200$. Equating this with Eq. (\ref{rhor}) and substituting Eq. (\ref{phidot}), we thus get the temperature of thermal bath as
\begin{equation}
T=\left(\frac{1}{2\pi C_R} \frac{Q}{(1+Q)^2} \lambda \,\phi^2M_{\text{Pl}}^2\right)^\frac{1}{4},
\label{Tk}
\end{equation}
where $C_R\equiv\pi^2 g_*/30$.
The two different phases in composite dissipation model differ in the dynamics of the dissipation parameter, which we discuss next.
	\subsection{Evolution of dissipation parameter in Phase-I:  $\Upsilon=C_1 T^3/M_{\text{Pl}}^2$ }
    We know, by definition $\Upsilon\equiv Q\cdot3H =C_{1} \frac{T^3}{M_{\text{Pl}}^2}$ in Phase-I. From this, we find $$T=\left(Q\cdot3H\cdot\frac{M_{\text{Pl}}^2}{C_1}\right)^\frac{1}{3},$$
    which can be equated with Eq. (\ref{Tk}) to get a relation between $Q$ and $\phi$ as,
    \begin{equation}
        \frac{\phi}{M_{\text{Pl}}}=\frac{C_1^2 \lambda^{1/2}}{24 \pi (2\pi C_R)^{3/2} Q^{1/2}(1+Q)^3}.
        \label{phi1}
    \end{equation}
    Taking the logarithm of both sides and  differentiating  with respect to the number of e-folds $N\equiv\text{ln}(a_e/a)$, and using the product rule $\frac{d\phi}{dN}=\frac{d\phi}{dt}\frac{dt}{dN}$ along with $dN/dt=-H$ (since 
$N$ is counted backward from the end of inflation), we obtain the expression for the evolution of
    dissipation parameter as:
	\begin{equation}
		\frac{dQ}{dN}= -\frac{4608\pi^4 C_R^3 Q^2 (1 + Q)^6}{\lambda C_1^4 (1 + 7 Q)}.
	\end{equation}
    This expression shows that as inflation proceeds ($N$ decreases), $Q$ increases. 
Starting from a chosen value of the dissipation parameter at the pivot scale $Q_P$, its value at any later stage of inflation can be determined by integrating the above equation.  

With these expressions, we can completely parameterize the dynamics of Phase-I of warm inflation in terms of the evolution of the dissipation parameter. The next step is to obtain the criteria that determine the end of this phase of inflation.\\

\textbf{Condition for the end of Phase-I:}  
We demand that the slow-roll parameter at the end of Phase-I obeys the condition
\begin{equation}
    \eta_{\phi,e1} = \frac{12}{8\pi}\,\frac{M_{\text{Pl}}^2}{\phi_{e1}^2} = x\,(1+Q_{e1}),
    \label{eta1}
\end{equation}
where $x < 1$ is a chosen constant.  The subscript $e1$ signifies the end values of parameters at the end of Phase-I of inflation.
We can also write this equation in terms of a parameter  
\begin{equation}
 \eta_{H,e1}=x
 \label{etae}
\end{equation}
where $\eta_{H,e1}\equiv\frac{\eta_{\phi,e1}}{1+Q_{e1}}.$
Substituting the expression for $\phi$ from Eq.~(\ref{phi1}) into Eq.~(\ref{eta1}) yields:
\begin{equation}
    Q_{e1} \,(1 + Q_{e1})^5 = x \,
    \frac{\lambda\, C_1^4}{6912\, \pi^4\, C_R^3}.
    \label{Qe}
\end{equation}
Solving this equation, we get the dissipation parameter value at the end of Phase-I, $Q_{e1}$.

	\subsection{Evolution of dissipation parameter in Phase-II: $\Upsilon= C_2 T^3/\phi^2$}
	In this phase, $\Upsilon \equiv Q\cdot3H=C_2 T^3/\phi^2$, which gives $$T=\left(Q\cdot3H\cdot\frac{\phi^2}{C_2}\right)^\frac{1}{3},$$
    which can be again equated with Eq. (\ref{Tk}) to get a relation between $Q$ and $\phi$ as,
    \begin{equation}
       \frac{\phi}{M_{\text{Pl}}}=\sqrt\frac{1}{8\pi}\left(\frac{64 C_2^4\lambda}{9 C_R^3}\frac{1}{Q(1+Q)^6}\right)^{\frac{1}{10}}.
        \label{phi2}
    \end{equation}
    Taking the logarithm of both sides, and following a similar procedure as above, we obtain the expression for the evolution of the dissipation parameter as:
	\begin{equation}
		\frac{dQ}{dN}= -40\left(\frac{9~ C_R^3}{64 ~\lambda ~C_2^4 }\right)^{1/5}\frac{Q^{6/5} (1 + Q)^{6/5}}{(1 + 7Q)}.
        \label{dQ2}
	\end{equation}
    In this expression, we see that $Q$ increases as inflation proceeds. 
    The value of $Q$ at the end of Phase-I, obtained in the previous subsection, is used as the initial condition for Phase-II. This value is then evolved for the required number of e-folds using the above equation. The slow-roll parameter in Phase-II satisfies the bound
\begin{equation}
    x\,(1+Q_{e1}) \ \leq\ \eta_{\phi} \ \leq\ (1+Q_{e2}),
\end{equation}
where $0 < x < 1$ is the same constant introduced in Phase-I, and $Q_{e2}$ refers to the value of the dissipation parameter at the end of Phase-II.\\

\textbf{Condition for the end of Phase-II:}  
The end of Phase-II as well as the inflationary era is determined by imposing the condition on the slow-roll parameter
\begin{equation}
    \eta_{\phi,e2} = \frac{12}{8\pi}\,\frac{M_{\text{Pl}}^2}{\phi_{e2}^2} = (1+Q_{e2}).
\end{equation}
This is equivalent to the condition on parameter $\eta_{H,e2}=1$, where $\eta_{H,e2}\equiv \frac{\eta_{\phi,e2}}{1+Q_{e2}}.$
Substituting the expression for $\phi$ from Eq.~(\ref{phi2}) into the above condition yields:
\begin{equation}
    Q_{e2} \,(1 + Q_{e2}) = \frac{1}{12^5}
    \frac{64~ C_2^4 ~\lambda}{9 ~C_R^3}.
    \label{Qe2}
\end{equation}
Solving Eq.~(\ref{Qe2}) gives the end value of the dissipation parameter $Q_{e2}$ at the end of inflation.
   
	\subsection{Numerical Framework}
We now discuss the methodology followed for constructing the Mathematica code for our CDWI model.\\

\textbf{Phase-I: Large-Scale Dynamics}\\
	
	The initial phase $\Upsilon=C_1 T^3/M_{\text{Pl}}^2$ is implemented through the following computational procedure:
	\begin{enumerate}
		\item \textbf{Normalization Constraint}: 
        We parameterize the primordial power spectrum using Eqs. (\ref{Hubble})-(\ref{phi1}) in terms of variables, $Q$,  $\lambda$ and $C_1$.
		Then, by fixing the dissipation strength $Q_P$ at the pivot scale and imposing the normalization of the scalar power spectrum at the pivot scale, $A_s=2.13\times 10^{-9},$ we obtain a relationship between $\lambda$ and $C_1$.
		
		\item \textbf{Phase Termination}: 
		The first phase concludes when the slow-roll parameter reaches a chosen threshold:
		$	\eta_{H,{e1}} = x. $
			This condition gives the dissipation strength $Q_{e1}$ at the end of Phase-I, in terms of $\lambda$ and $C_1$, as given in Eq. (\ref{Qe}).
		
		\item \textbf{Numerical Integration}:
        We integrate $dQ/dN$ from the pivot scale to the end of Phase-I, with the limits as a fixed $Q_P$, $N_P=N_1$ (fixed), $N_e=0$, and $Q_{e1}$ (obtained as above). This gives another relation between  $\lambda$ and $C_1$. Solving this together with the one obtained from the normalization condition,  we determine both $\lambda$ and $C_1$.

		\item \textbf{Dynamical Evolution}: Once $\lambda$ and $C_1$ are known, we can determine $Q(N)$ by solving $dQ/dN$ and thus
		the full system of dynamical variables $\{\phi(N), T(N), H(N)\}$ can be evaluated for $N_1$ e-folds of  Phase-I.
	\end{enumerate}

	As the inflaton rolls down the potential, the second term in the composite dissipation increases. 
    The relative growth of the second dissipation term controls the transition between phases:
	\begin{equation}
		C_2 = C_1 \left(\phi_{e1}/M_{\text{Pl}}\right)^2,    
	\end{equation}
	where $\phi_{e1}$ is the field value at the end of Phase-I. This ensures a continuous transition where the second term becomes dominant precisely after $N_1$ e-folds.\\

   \textbf{Phase-II: Small-Scale Dynamics}\\
	
	The second inflationary phase driven by $\Upsilon=C_2 T^3/\phi^2$ is characterized by:
	\begin{itemize}
		\item The initial values of parameters $Q, T, H$ are taken as obtained at the endpoint of Phase-I.
		\item The inflaton self-coupling $\lambda$ is taken to be the same during both Phase-I and II. 
		\item  The slow roll parameter evolves from value  $x\,(1+Q_{e1}) \ \leq\ \eta_{\phi} \ \leq\ (1+Q_{e2})$  in Phase-II. As we know $\lambda$ and $C_2$, we can solve Eq. (\ref{Qe2}) to obtain the dissipation parameter at the end, $Q_{e2}$. Then, using Eq. (\ref{dQ2}), we can
       determine the $N_2$ number of e-folds of Phase-II. 
	\end{itemize}
	
	The complete inflationary epoch is made to last for approximately 50 e-folds ($N_1 + N_2 \approx 50$), by carefully selecting the model parameters along with
   maintaining consistency with the CMB observational constraints.

	\section{Results}
    \label{results}
    In this Section, we analyze the impact of different choices of model parameters in the CDWI model on the primordial power spectrum.

    \subsection{Effect of number of e-folds of Phase-I}
First, we examine the effect of varying the duration of Phase-I by considering different values for the number of e-folds $N_1$. 
The results are shown in Table \ref{table1}, \ref{table2}, and \ref{table3} for $N_1=20, 25,$ and $30$ respectively. 
We list the results for different values of the transition parameter $\eta_{H,e1}=x$, defined in Eq. (\ref{etae}).
We have considered a $Q_P$ value of 100 in generating these Tables. 
We observe that for a given value of $x$ (for example, 1/4), on increasing $N_1$ from 20 to 30, the value of the dissipation parameter $Q_{e1}$ at the end of Phase-I increases. This is expected to happen since an extended duration of inflation ($N_1$) leads to an extended evolution of the dissipation parameter $Q$ from integration of $dQ/dN$. 

Further, the primordial power spectrum becomes flatter (i.e., $|n_s-1|$ decreases) for larger $N_1$ values. This suggests that for $N_1\geq 30$, the spectral index can be favorable with the recent results from \textit{Atacama Cosmology Telescope} (ACT), $n_s=0.9743\pm 0.0034$ \cite{ACT:2025fju}.
We also observe that the tensor-to-scalar ratio in this CDWI model is very suppressed $\mathcal{O}(10^{-18})$. This demonstrates that Phase-I of CDWI can allow a strong dissipative regime and large $N_1$ compatible with the CMB observations.
Note that we cannot keep on increasing $N_1$, as it would also increase the total duration of inflation ($N_1+N_2)$.

    A key finding in this analysis is that for a fixed value of $x$, the duration of Phase-II ($N_2$) remains essentially unchanged across all these tables. This implies that the transition parameter $x$ is solely responsible for determining $N_2$, for example:
\begin{itemize}
    \item $x=1/4$ yields $N_2 \approx 15.7$
    \item $x=1/6$ gives $N_2 \approx 26$
    \item $x=1/8$ results in $N_2 \approx 36.7$.
\end{itemize}
As $x$ decreases, the number of e-folds of Phase-II increases.
This relationship indicates that once $x$ is specified, the value of $N_2$ is automatically determined, and thus, $N_1$ must then be chosen such that the total inflationary duration $N_1 + N_2$ falls within the observationally favored range of 50-60 e-folds.

        \begin{table}[H] 
		\centering
		\begin{tabular}{|l|c|c|c|}
			\hline
			$\eta_{H, e1}=x$ in Phase-I & {$x = 1/4$}    & {$x = 1/6$}    & {$x= 1/8$}   \\  \hline
			End value $Q_{e1}$ in Phase-I 	& {125.45}  & {119.62}  &{116.08} \\ \hline
			Spectral Index $n_s$	& {0.9544}  & {0.9599}  & {0.9641}  \\ \hline
			Tensor-to-scalar ratio $r$	& {$2.46\times 10^{-18}$} &  {$2.18\times10^{-18}$} & {$1.96\times10^{-18}$}  \\ \hline
			Number of e-folds in Phase-II, $N_2$	& {15.72} & {26.23} & {36.71} \\ \hline
		\end{tabular} 
		\caption{Different parameters obtained by considering $N_1=20$ e-folds in Phase-I  for various values of $x$ 
			in the condition $\eta_{H, e1}=x$. Here, the dissipation parameter value at the pivot scale is fixed as $Q_P=100$.}
         \label{table1}   
	\end{table}

        \begin{table}[H] 
		\centering
		\begin{tabular}{|l|c|c|c|}
			\hline
			$\eta_{H, e1}=x$ in Phase-I & {$x = 1/4$}    & {$x = 1/6$}    & {$x= 1/8$}   \\  \hline
			End value $Q_{e1}$ in Phase-I 	& {129.08}  & {122.72}  &{118.78} \\ \hline
			Spectral Index $n_s$	& {0.9618}  & {0.9657}  & {0.9689}  \\ \hline
			Tensor-to-scalar ratio $r$	& {$2.08\times 10^{-18}$} &  {$1.87\times10^{-18}$} & {$1.71\times10^{-18}$}  \\ \hline
			Number of e-folds in Phase-II, $N_2$	& {15.77} & {26.30} & {36.72} \\ \hline
		\end{tabular} 
		\caption{Different parameters obtained by considering $N_1=25$ e-folds in Phase-I  for various values of $x$. Here also $Q_P=100$.}
         \label{table2}   
	\end{table}

\begin{table}[H] 
		\centering
		\begin{tabular}{|l|c|c|c|}
			\hline
			$\eta_{H, e1}=x$ in Phase-I & {$x = 1/4$}    & {$x = 1/6$}    & {$x= 1/8$}   \\  \hline
			End value $Q_{e1}$ in Phase-I 	& {132.26}  & {125.45}  &{121.21} \\ \hline
			Spectral Index $n_s$	& {0.9672}  & {0.9701}  & {0.9725}  \\ \hline
			Tensor-to-scalar ratio $r$	& {$1.80\times 10^{-18}$} &  {$1.64\times10^{-18}$} & {$1.51\times10^{-18}$}  \\ \hline
			Number of e-folds in Phase-II, $N_2$	& {15.70} & {26.28} & {36.68} \\ \hline
		\end{tabular} 
		\caption{Different parameters obtained by considering $N_1=30$ e-folds in Phase-I  for various values of $x$. Here also $Q_P=100$.}
         \label{table3}   
	\end{table}
    
\subsection{Effect of transition parameter $\eta_{\phi,e1}$}
	\label{sec:x}
	Next, we discuss the effect of the parameter deciding the end of Phase-I, $\eta_{H,e1}=x$.
 For a fixed value of dissipation parameter at the pivot $Q_P$, we consider different values of $x$ and list the results in Table \ref{table1}, \ref{table2}, and \ref{table3} for $N_1=20, 25, 30,$ respectively. 
    
    In all these Tables, we see that as we decrease the value of $x$ from $1/4$ to $1/8$, the dissipation parameter at the end of Phase-I, $Q_{e1}$, decreases. This is because $x$ appears directly in the equation for obtaining $Q_{e1}$, see  Eq. (\ref{Qe}). It is to be noted that 
for fixed $N_1$ and $Q_P$, the parameters $\lambda$ and $C_1$ also vary with $x$, with $\lambda$ increasing and $C_1$ decreasing as $x$ increases. The relative variations in $\lambda$ and $C_1$ are, however, small compared to the change in $x$. Thus, the solution for $Q_{e1}$ in Eq. (\ref{Qe}) is predominantly determined by the variation of $x$, implying that a reduction in $x$ leads to a corresponding decrease in $Q_{e1}$.
     
   Further, it can be seen that for a given $N_1$, a smaller $x$ and corresponding $Q_{e1}$ leads to a flatter primordial power spectrum, i.e., a smaller $|n_s-1|$, which is favorable with the ACT results.
Also, as previously discussed, the tensor-to-scalar ratio in this CDWI model is very suppressed which hints that this model can be made compatible with the CMB observations.
	
We further see that for a given $N_1$, as we decrease the value of $x$, the number of e-folds $N_2$ increases, and this is true across all Tables. This behavior is expected since the slow roll parameter $\eta_\phi$ in Phase-II has to evolve from $x(1+Q_{e1})$ to $(1+Q_{e2})$, therefore, a smaller $x$ requires more number of e-folds in Phase-II to reach the end of inflation condition. 

We again stress that the transition factor $x$ solely controls the number of e-folds $N_2$. As the total number of e-folds of inflation should typically lie in between $50-60$, thus, we cannot reduce $x$ much below 1, otherwise the Phase-II will itself last for more than $50-60$ efolds and the presence of Phase-I will not be very evident on the CMB scale dynamics.

	\subsection{Effect of dissipation parameter  value at the pivot scale $Q_P$}
We now analyze the impact of different values of the dissipation parameter at the pivot scale during Phase-I of inflation. For this study, we consider the total number of e-folds of inflation $N_1+N_2=50$. Consequently, we choose two models, such that $N_1$ is sufficiently large and $x$ is small to have the spectral index compatible with the ACT results:
\begin{itemize}
    \item $x=1/4$ (yielding $N_2 \approx 15.7$) and $N_1 = 35$.
    \item $x=1/3$ (yielding $N_2 \approx 10.5$)
    and $N_1 = 40$.
\end{itemize}
 The resulting parameters for various $Q_P$ values are presented in Tables~\ref{table4}--\ref{table5}.
Our analysis reveals two key findings:
\begin{itemize}
    \item The spectral index is more for larger $Q_P$ values, suggesting that compatibility with ACT observations requires relatively large $Q_P$ values in Phase-I.
    \item The tensor-to-scalar ratio for $Q_P=100$ is lower by several orders of magnitude compared to $Q_P=10$.
\end{itemize}

Notably, we observe that $N_2$ remains constant for a given $x$ regardless of $Q_P$ variations. This reinforces our earlier conclusion in subsection~\ref{sec:x} that $N_2$ is exclusively determined by the parameter $x$.

\begin{table}[H] 
		\centering
		\begin{tabular}{|l|c|c|c|}
			\hline
			Dissipation parameter at pivot & $Q_P=100$   & $Q_P=50$  & $Q_P=10$   \\  \hline
			End value $Q_{e1}$ in Phase-I  	& {135.12}  & 67.85  & 13.77\\ \hline
			Spectral Index $n_s$	& {0.9712}  & {0.9677}   & 0.9607\\ \hline
			Tensor-to-scalar ratio $r$	& {$1.58\times 10^{-18}$} &  {$1.47\times 10^{-15}$} & {$5.19\times 10^{-10}$}\\ \hline
			Number of e-folds in Phase-II, $N_2$	& {15.78} & {15.72} & 15.60\\ \hline
		\end{tabular} 
		\caption{Different parameters obtained by considering $x=1/4~ (N_1=35)$  with $Q_P=100, 50,$ and $10$, respectively.}
         \label{table4}   
	\end{table}
    
  \begin{table}[H] 
		\centering
		\begin{tabular}{|l|c|c|c|}
			\hline
			Dissipation parameter at pivot & $Q_P=100$   & $Q_P=50$  & $Q_P=10$   \\  \hline
			End value $Q_{e1}$ in Phase-I  	& {143.56}  & 72.13  & 14.68\\ \hline
			Spectral Index $n_s$	& {0.9733}  & {0.9701}   & 0.9636\\ \hline
			Tensor-to-scalar ratio $r$	& {$1.47\times 10^{-18}$} &  {$1.37\times 10^{-15}$} & {$4.82\times 10^{-10}$}\\ \hline
			Number of e-folds in Phase-II, $N_2$	& {10.47} & {10.48} & 10.40\\ \hline
		\end{tabular} 
		\caption{Different parameters obtained by considering $x=1/3 ~(N_1=40)$  with $Q_P=100, 50,$ and $10$, respectively.}
         \label{table5}   
	\end{table}

	\subsection{Effect of the Growth function}
\begin{figure}
    \includegraphics[width=0.45\textwidth]{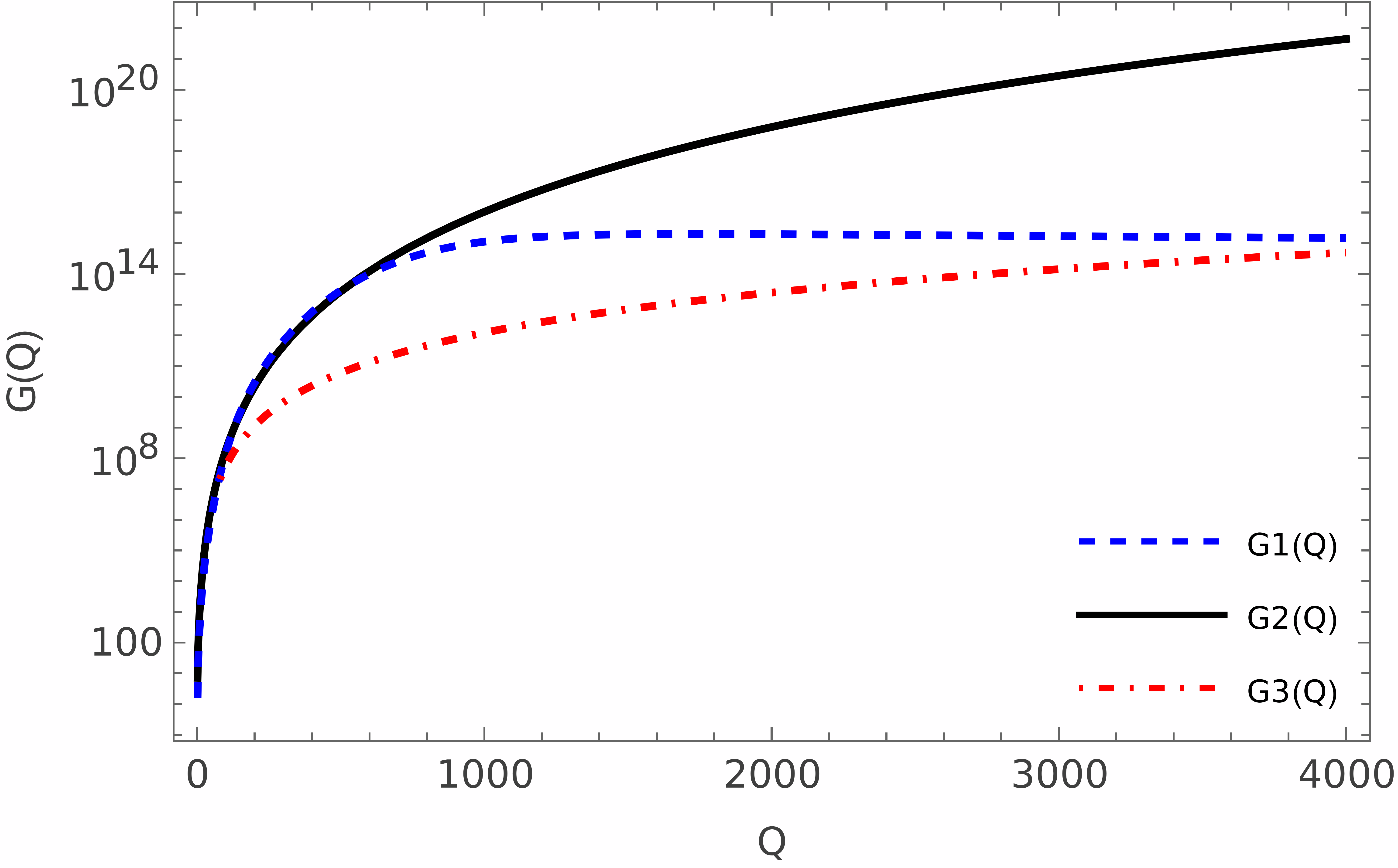}
    \caption{Plots for $G_1(Q)$, $G_2(Q)$, and $G_3(Q)$, in dashed blue, solid black, and dot-dashed red, respectively. Note the logarithmic scale on the vertical axis. One can see that they agree at low values of $Q$, but they differ by many orders of magnitude at intermediate and/or higher values, probed in the current work.}
    \label{G123}
    \end{figure}

    In this subsection, we investigate how different parameterizations of the growth function $G(Q)$ affect the primordial power spectrum. The growth factor is determined numerically by solving the coupled differential equations governing inflaton and radiation fluctuations. We examine three distinct forms of $G(Q)$ that arise in the literature \cite{Benetti:2016jhf,Das:2020xmh,Montefalcone:2023pvh} for dissipation coefficients with cubic temperature dependence ($\Gamma \propto T^3$):
	
	\begin{align}
		G_1(Q)&=\frac{1+6.12 ~Q^{2.73}}{(1+6.96 ~Q^{0.78})^{0.72}} + \frac{0.01 Q^{4.61}(1+4.82\times10^{-6}~ Q^{3.12})}{(1+6.83\times10^{-13} ~Q^{4.12})^2},\\
		G_2(Q)&=	10^{\sum_{n=1}^4 a_n y^n}      \hspace{1.2cm}       (\textrm{Logarithmic fitting)},\\
		G_3(Q)&=1+4.981~ Q^{1.946}+ 0.17 ~Q^{4.330}.
	\end{align}
	where $y\equiv \textrm{log}_{10}(1+Q)$,~ $a_1=2.86,~ a_2=-0.18,~ a_3=0.55,~ a_4=-7.14\times10^{-2}$.
One can observe the behavior of these functions in Fig.~\ref{G123}. It can be seen that the fitting functions match extremely well for
$Q<100$, but differ by many orders of magnitude for larger values of the dissipation strength. 
$G_3(Q)$ was used in Refs. \cite{Benetti:2016jhf,Motaharfar:2018zyb} for small value of dissipation parameter $Q<100$, while $G_1(Q)$ was derived with more free parameters compared to simpler form of $G_3(Q)$, in Ref.~\cite{ Das:2020xmh} for wider range $Q<10^3$.
The logarithmic function $G_2(Q)$ proposed in recent Ref.~\cite{Montefalcone:2023pvh} fits the data well over the entire range of $Q \in[10^{-7},10^4]$, probed in the current work. 

	As discussed above, we consider two case studies:
 $N_1 = 35$ for $x = 1/4$ ($N_2 \approx 15.7$), and
   $N_1 = 40$ for $x = 1/3$ ($N_2 \approx 10.5$).
    In Tables \ref{table7}-\ref{table8}, we list the obtained parameters using the growth functions $G_1(Q)$, $G_2(Q)$ and $G_3(Q)$, respectively. Our analysis reveals that while the three growth functions produce similar values for parameters $Q_{e1}$ and $N_2$, they yield distinct values for the spectral tilt of the primordial power spectrum. The function $G_1(Q)$ yields larger $n_s$ value (flatter primordial power spectrum), while $G_3(Q)$ gives smaller $n_s$ values (steeper power spectrum). 

\begin{table}[H] 
		\centering
		\begin{tabular}{|l|c|c|c|}
			\hline Form of growth factor taken
			 & $G_1(Q)$  & $G_2(Q)$ & $G_3(Q)$ \\  \hline
			End value $Q_{e1}$ in Phase-I 	& {135.11}  & {135.12}  & 135.11\\ \hline
			Spectral Index $n_s$	& {0.9753}  & {0.9712}   & 0.9626\\ \hline
			Tensor-to-scalar ratio $r$	& {$2.11\times 10^{-18}$} &  {$1.58\times10^{-18}$}  & {$7.85\times 10^{-18}$}\\ \hline
			Number of e-folds in Phase-II, $N_2$	& {15.80} & {15.78} & 15.68\\ \hline
		\end{tabular} 
		\caption{Different parameters obtained by considering $x=1/4~ (N_1=35)$ with different forms of growth function $G_1(Q)$, $G_2(Q)$, and $G_3(Q)$. Here we have considered $ Q_P=100$.}
         \label{table7}   
	\end{table}

\begin{table}[H] 
		\centering
		\begin{tabular}{|l|c|c|c|}
			\hline Form of growth factor taken
			 & $G_1(Q)$  & $G_2(Q)$ & $G_3(Q)$ \\  \hline
			End value $Q_{e1}$ in Phase-I 	& {143.57}  & {143.56}  & 143.57\\ \hline
			Spectral Index $n_s$	& {0.9771}  & {0.9733}   & 0.9654\\ \hline
			Tensor-to-scalar ratio $r$	& {$1.96\times 10^{-18}$} &  {$1.47\times10^{-18}$}  & {$7.29\times 10^{-18}$}\\ \hline
			Number of e-folds in Phase-II, $N_2$	& {10.53} & {10.47} & 10.53\\ \hline
		\end{tabular} 
		\caption{Different parameters obtained by considering $x=1/3 ~(N_1=40)$ with different forms of growth function $G_1(Q)$, $G_2(Q)$, and $G_3(Q)$. Here we have considered $ Q_P=100$.}
         \label{table8}   
	\end{table}

\begin{figure}[H]
		\centering
		\subfigure[]{
			\label{fig:1a}
			\includegraphics[width=0.45\textwidth]{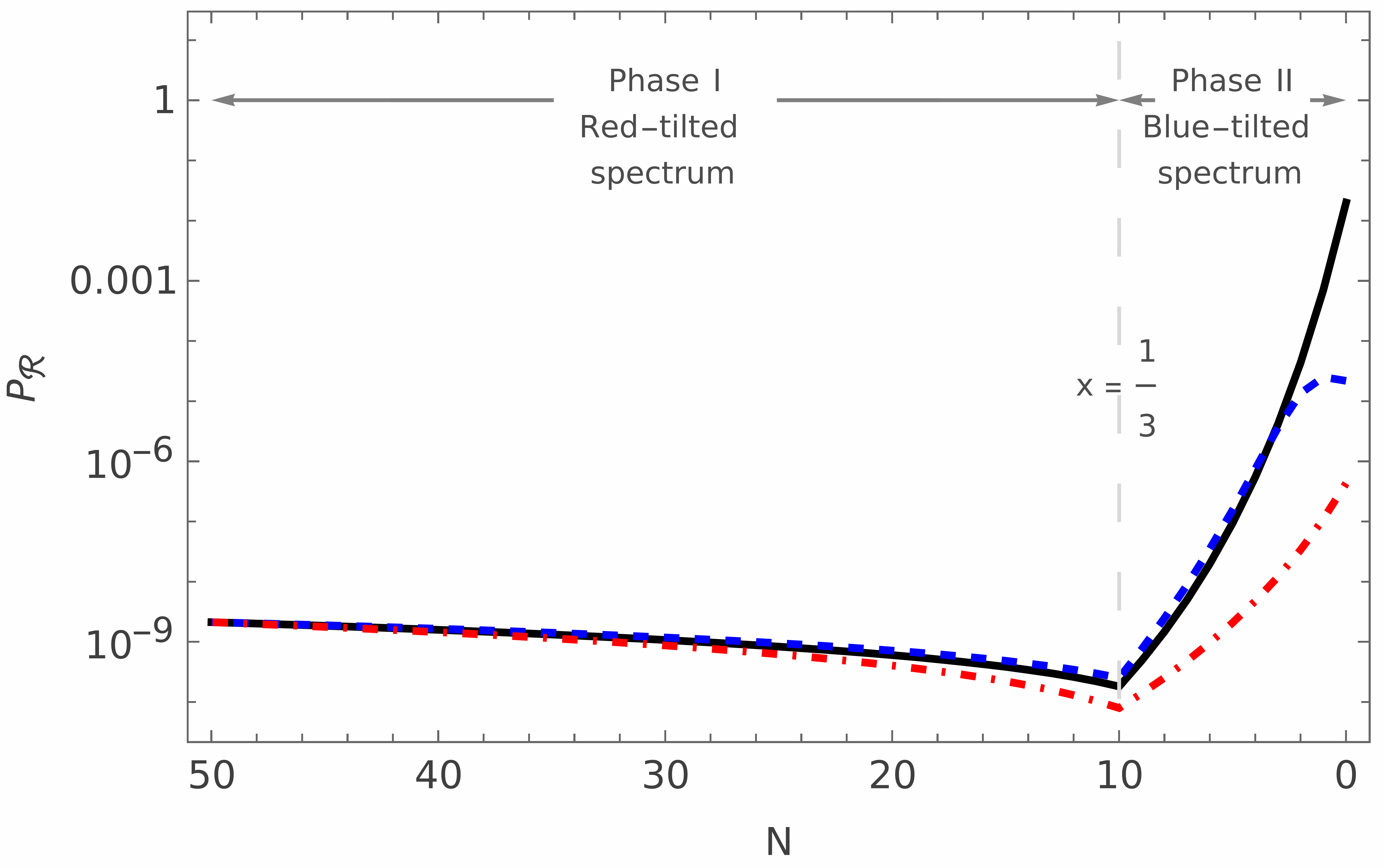}}
		\hspace{0.5cm}
		\subfigure[]{
			\label{fig:1b}
			\includegraphics[width=0.45\textwidth]{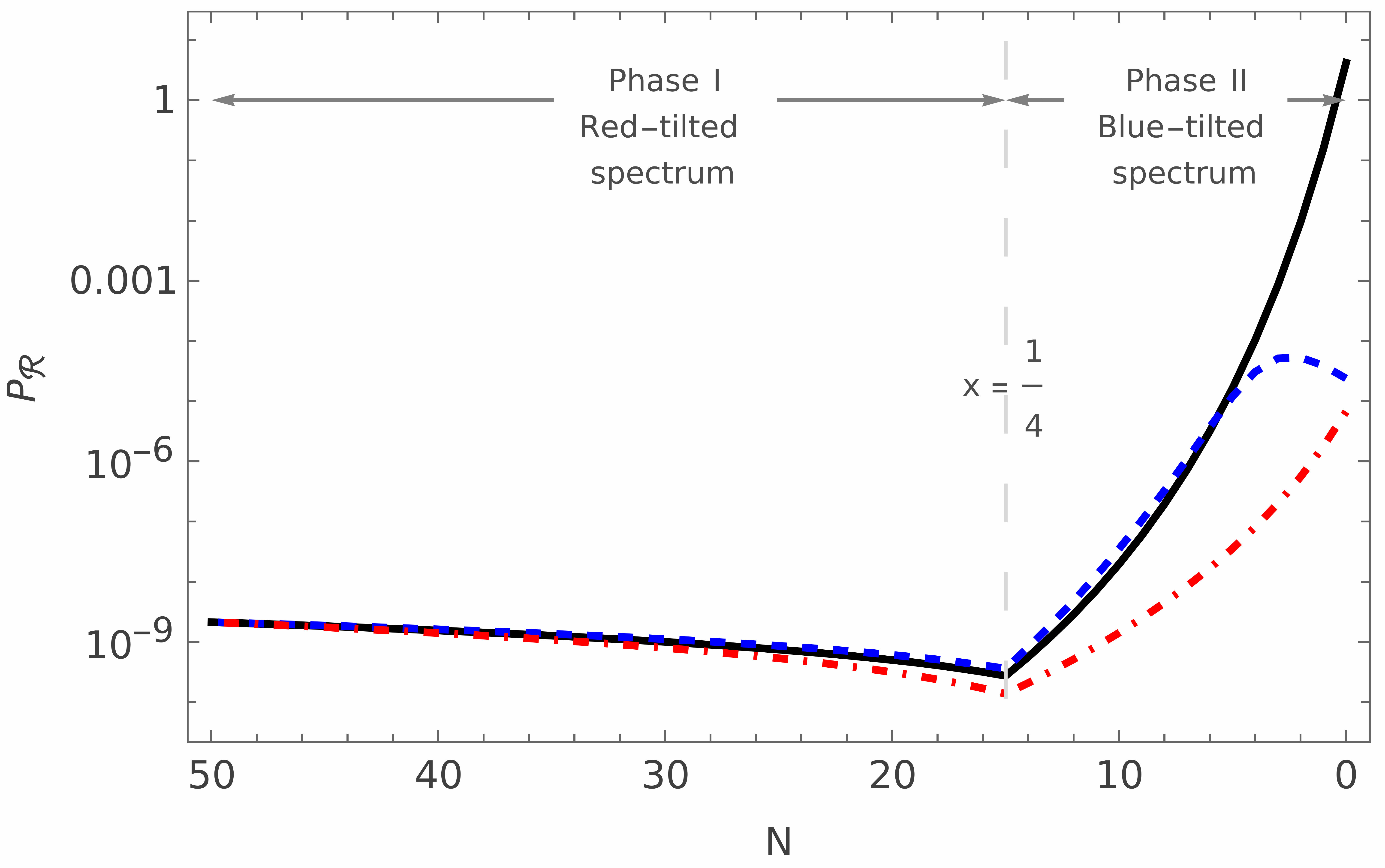}}	
		\caption{Primordial power spectrum as a function of number of e-folds for our CDWI models. The pivot is chosen at $N_P=50$ and the value of the dissipation parameter $Q_P=100$. The two cases considered are: (a) Phase-I lasts for 40 e-folds and Phase-II for 10 e-folds ($x=1/3$), and (b) Phase-I lasts for 35 e-folds and Phase-II for 15 e-folds ($x=1/4$). The dashed blue, solid black, and dot-dashed red lines represent the spectrum with growth factor $G_1(Q)$, $G_2(Q)$, and $G_3(Q)$, respectively.}
		\label{Pk} 	
	\end{figure}

    In Fig. \ref{Pk}, we plot the evolution of the primordial power spectrum in our CDWI model for the total $50$ e-folds of inflation. 
    Our findings show that:
    \begin{itemize}
        \item With $G_1(Q)$, the primordial power spectrum is red-tilted in Phase-I with a flatter tilt and transitions to a blue-tilt in Phase-II, as shown by the dashed blue line. While the spectrum reaches a maximum amplitude of $\mathcal{O}(10^{-5})$, this enhancement remains insufficient for primordial black hole (PBH) formation.
        \item 
        The $G_2(Q)$ case, represented by the black solid line, is of particular interest as it exhibits the strongest growth of the power spectrum at small scales, thus providing viable conditions for PBH formation.
    \item For the two case studies examined in this work, it is found that $G_3(Q)$, represented with a dot-dashed red line, yields a spectral index inconsistent with recent CMB measurements. Consequently, $G_3(Q)$ is not pursued further, as $n_s$ predictions conflict with ACT observations.
    
    \end{itemize}

	\section{Background evolution of CDWI models and Swampland conjecture}
    \label{back}
\begin{figure}[]
		\centering
		\subfigure[]{
			\label{fig:3a}
			\includegraphics[width=0.45\textwidth]{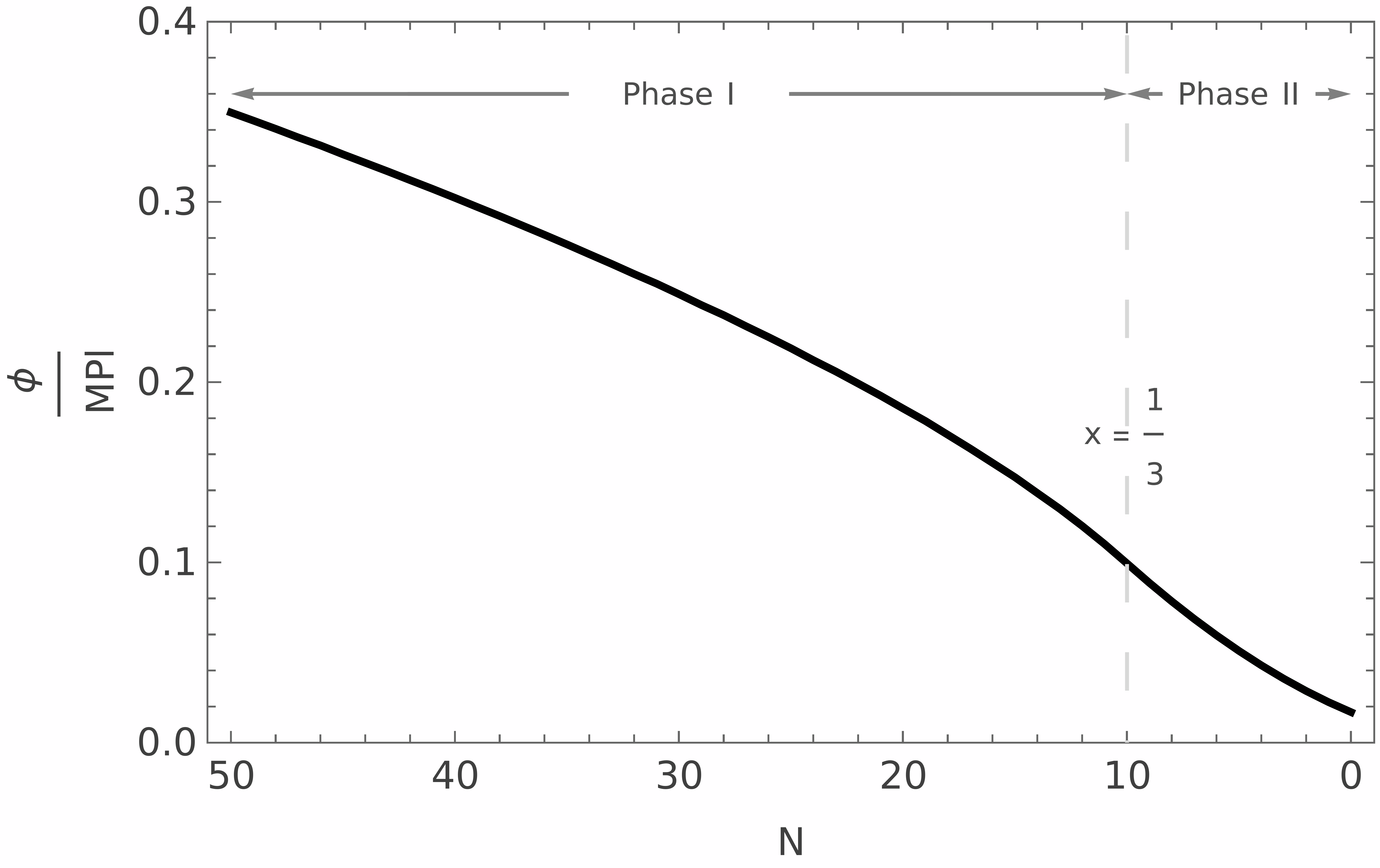}}
		\hspace{0.5cm}
		\subfigure[]{
			\label{fig:3b}
			\includegraphics[width=0.45\textwidth]{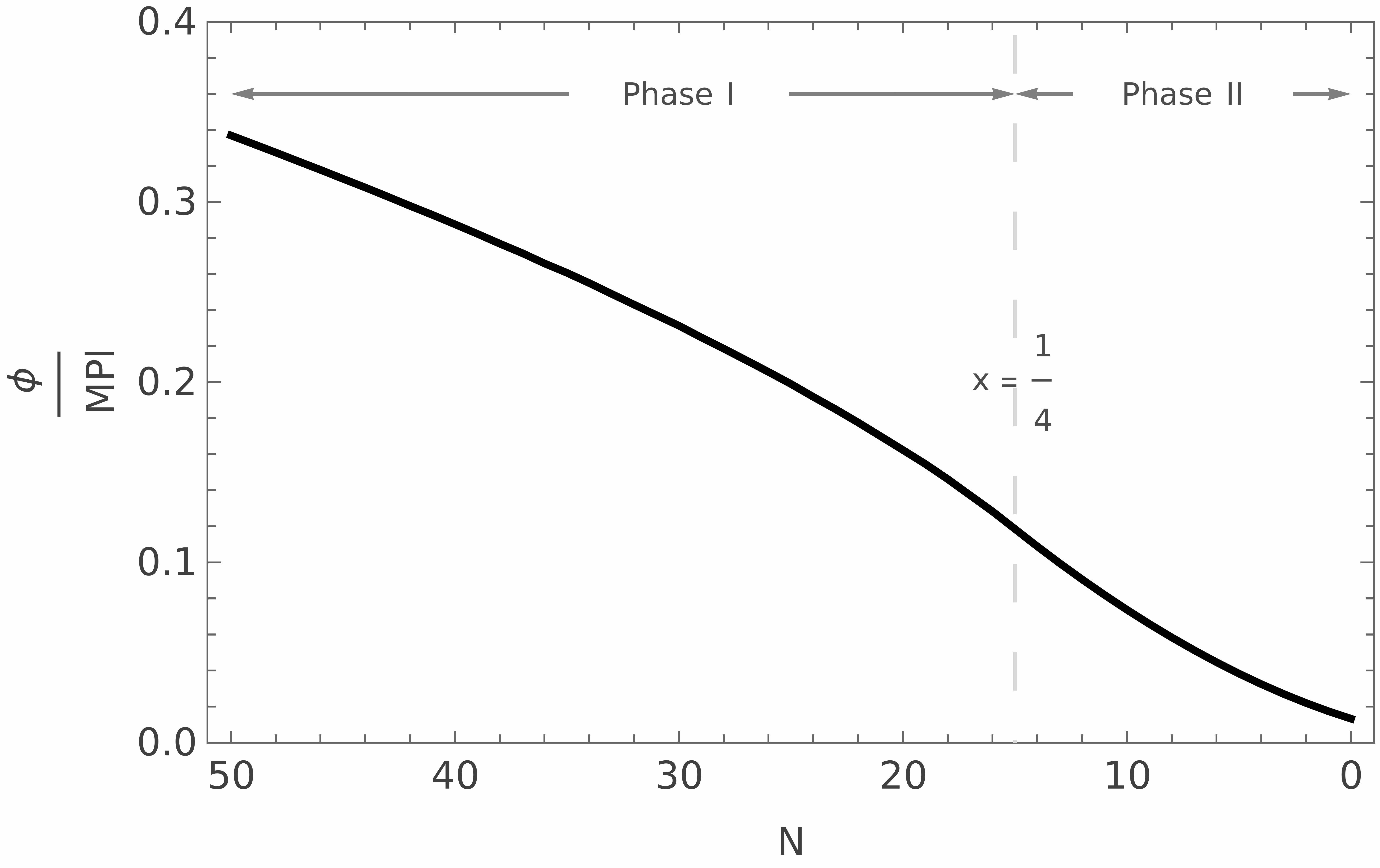}}	
		\caption{Plots of inflaton field evolution with number of e-folds for the composite dissipation models. (a) Phase-I lasts for 40 e-folds and Phase-II for 10 e-folds ($x=1/3$). (b) Phase-I lasts for 35 e-folds and Phase-II for 15 e-folds ($x=1/4$). }
		\label{phi} 	
	\end{figure}

\begin{figure}[H]
		\centering
		\subfigure[]{
			\label{fig:4a}
			\includegraphics[width=0.45\textwidth]{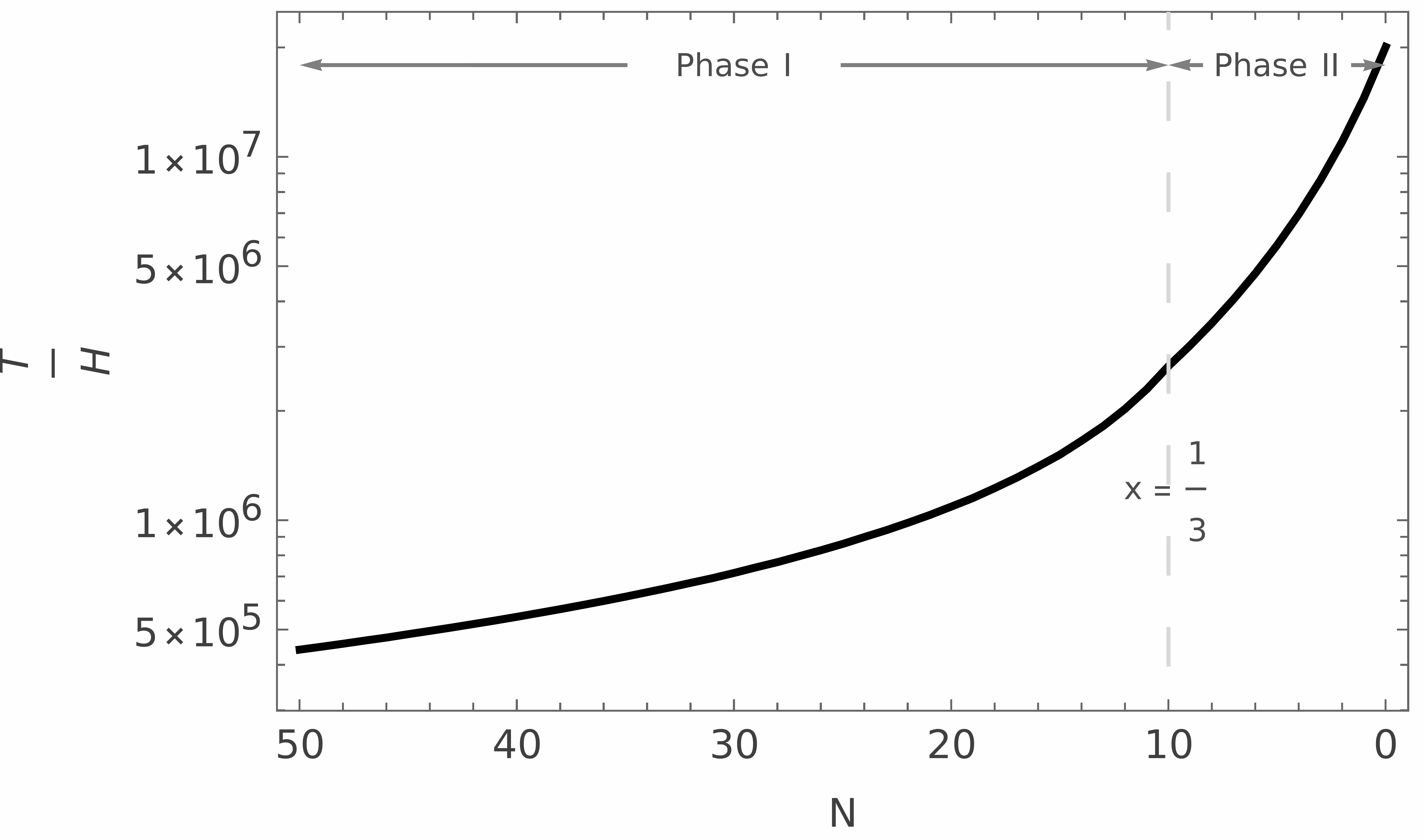}}
		\hspace{0.5cm}
		\subfigure[]{
			\label{fig:4b}
			\includegraphics[width=0.45\textwidth]{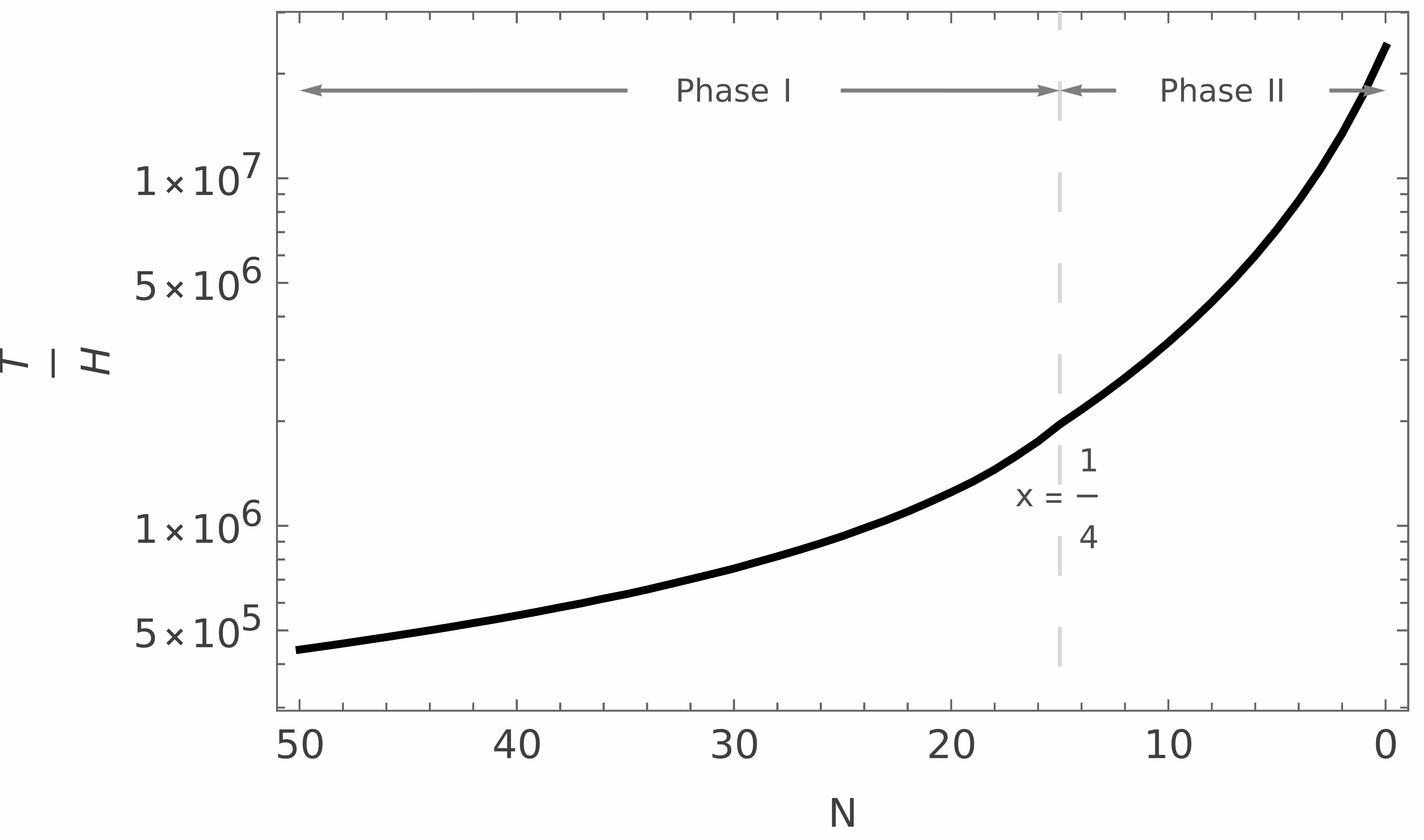}}	
		\caption{Plots of $T/H$ with the number of e-folds for the composite dissipation model. (a) Phase-I lasts for 40 e-folds and Phase-II for 10 e-folds ($x=1/3$). (b) Phase-I lasts for 35 e-folds and Phase-II for 15 e-folds ($x=1/4$).}
		\label{TbyH} 	
	\end{figure}

\begin{figure}[H]
		\centering
		\subfigure[]{
			\label{fig:5a}
			\includegraphics[width=0.45\textwidth]{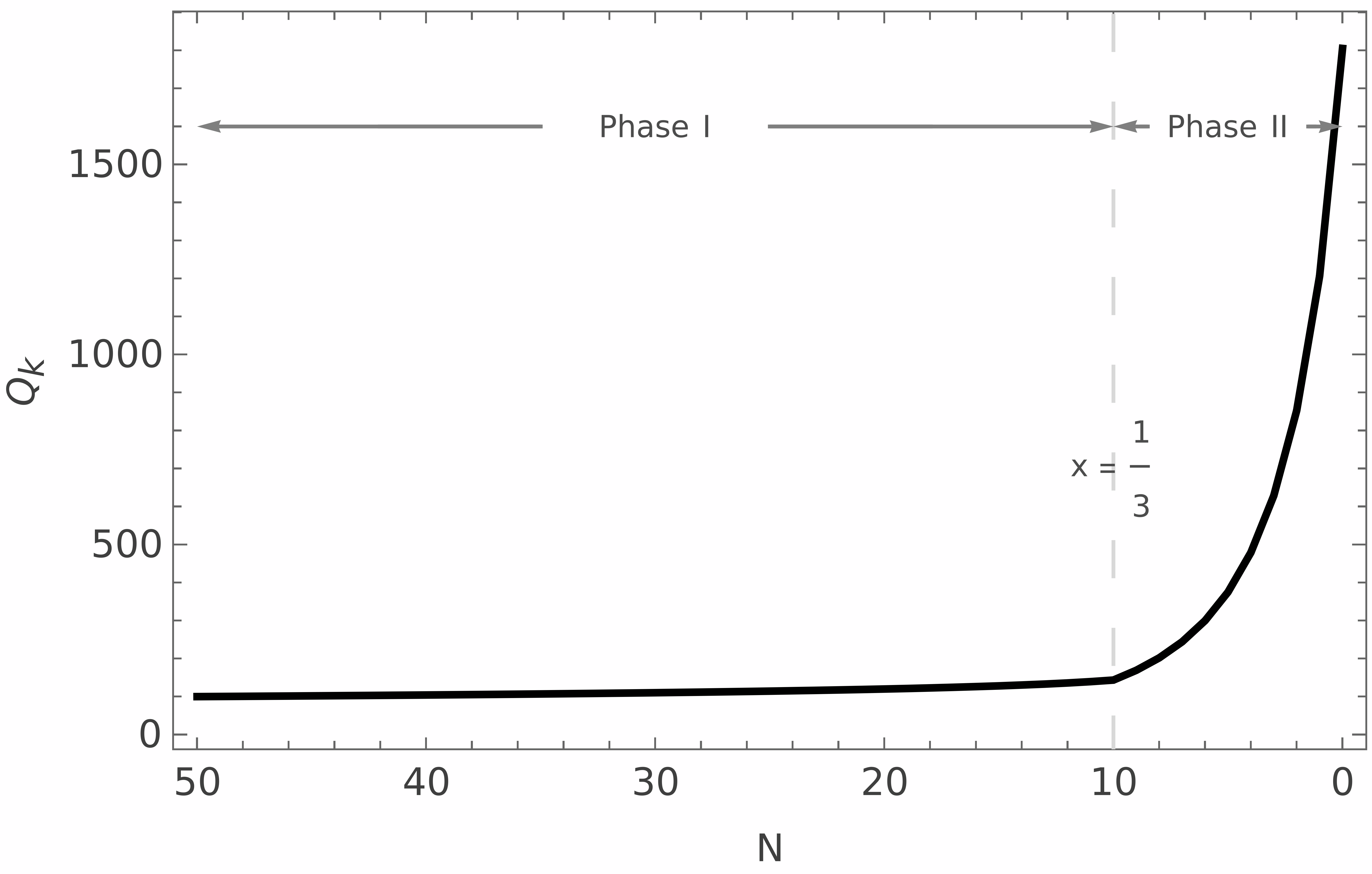}}
		\hspace{0.5cm}
		\subfigure[]{
			\label{fig:5b}
			\includegraphics[width=0.45\textwidth]{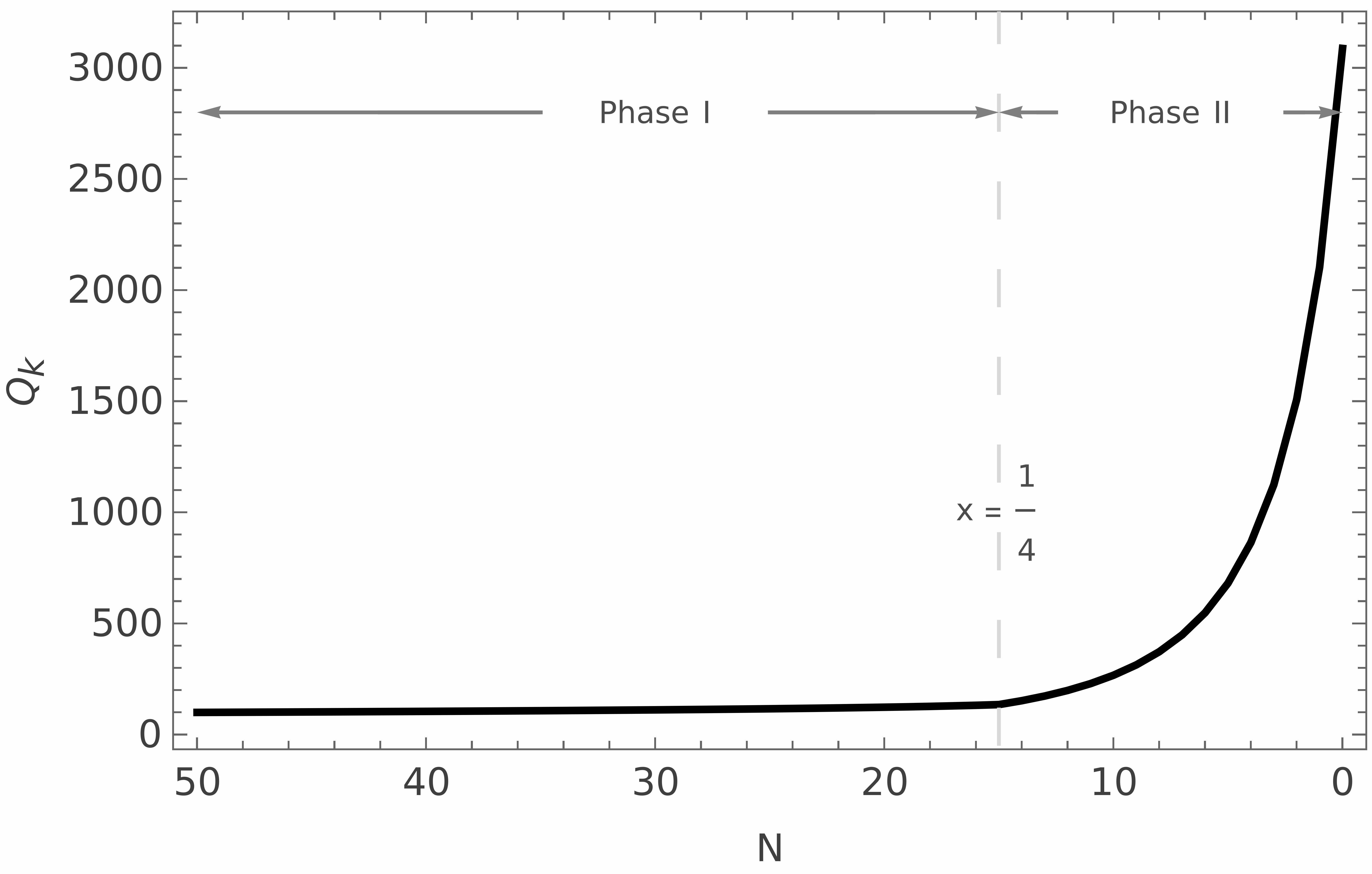}}	
		\caption{Plots of the evolution of dissipation parameter $Q(N)$ versus the number of e-folds for the composite dissipation model. Here we consider that the combined number of e-folds are $50$ and at the pivot $N_P=50$, $Q_P=100$. (a)  Phase-I lasts for 40 e-folds and Phase-II for 10 e-folds ($x=1/3$), and (b) Phase-I lasts for 35 e-folds and Phase-II for 15 e-folds ($x=1/4$).}
		\label{Q} 	
	\end{figure}

	We now show the evolution of parameters $\phi/M_{Pl}$, $T/H$, and $Q(N)$ for our CDWI models in Figures \ref{phi}, \ref{TbyH}, and \ref{Q}, respectively.  We consider a total of $50$ e-folds of inflation with composite dissipation and $Q_P=100$ at the pivot scale. As discussed in the previous section, we identify two models (a) Phase-I lasts for 40 e-folds and Phase-II for 10 e-folds ($x=1/3$), and  (b) Phase-I lasts for 35 e-folds and Phase-II for 15 e-folds ($x=1/4$). In these models, the spectral index of the primordial power spectrum is compatible with the ACT results on the CMB scales, as shown in Tables \ref{table4} and \ref{table5}.  
    
    We plot the evolution of inflaton field with the number of e-folds in Fig. \ref{phi}.
    We see from Fig. \ref{fig:3a} and \ref{fig:3b} that the inflaton field rolls down the potential during $N=50$ to $N=10 ~(15)$ in Phase-I and then from $N=10 ~(15)$ to the end of inflation $N=0$ in Phase-II. Throughout the evolution, the slow-roll conditions are obeyed, and the transition from Phase-I to Phase-II is governed by the parameter $\eta_{H,e1}=x$. In these models with quartic potential, as $\eta_H$ is more than $\epsilon_H$, maintaining the consistency of the $\eta_H$ parameter also ensures the slow-roll validity of $\epsilon_H$. 

   In Tables \ref{table8} and \ref{table9}, we list the values of parameters $\lambda, C_1, C_2$ obtained for the two models of CDWI considered. We 
also show the value of field excursion $\Delta \phi/ M_\text{Pl}$ and scale of inflation $V^{\frac{1}{4}}$ in these Tables. It is to be seen that inflaton field has a sub-Planckian excursion, which satisfies the swampland distance conjecture, $|\Delta \phi|/ M_\text{Pl} \leq a$, where the constant $a\sim \mathcal{O}(1)$. However, the trans-Planckian conjecture, which requires energy scale of inflation $V^{\frac{1}{4}}<10^9$ GeV, is not obeyed in these models.
\begin{table}[H]
		\centering	
		\begin{tabular}{|c|c|c|c|c|c|}
			\hline
			Parameter	& $\lambda$ & $C_1$ & $C_2$ & $\Delta \phi/ M_\text{Pl}$ & $V^{\frac{1}{4}} \text{(GeV)}$ \\ \hline
			Value	& $4.90\times10^{-27}$ & $5.71\times10^{12}$ & $5.65\times10^{10}$ & $0.33$ & $1.11\times10^{12}$ \\ \hline
		\end{tabular}
		\caption{Values of background parameters obtained for CDWI model with $N_1=40$ and $N_2=10~(x=1/3) $. Here $Q_P=100.$}
		\label{table8}
	\end{table}

	\begin{table}[H]
	\centering	
	\begin{tabular}{|c|c|c|c|c|c|}
		\hline
		Parameter	& $\lambda$ & $C_1$ & $C_2$ & $\Delta \phi/ M_\text{Pl}$  & $V^{\frac{1}{4}}  \text{(GeV)}$  \\ \hline
		Value	& $6.12\times10^{-27}$ & $5.30\times10^{12}$ & $7.46\times10^{10}$ & $0.32$ & $2.42\times10^{12}$ \\ \hline
	\end{tabular}
	\caption{Values of background parameters obtained for CDWI model with $N_1=35$ and $N_2=15~  (x=1/4)$. Here $Q_P=100.$}
	\label{table9}
\end{table}

	We next plot the evolution of $T/H$ as a function of the number of e-folds, for the two CDWI models in 
  Fig. \ref{fig:4a} and \ref{fig:4b}, respectively. We see that the factor $T/H$ is always greater than 1, which is required to maintain warm inflation. Further, this factor increases during both phases of inflation, thereby maintaining the consistency of warm inflation throughout.  
  
In Fig. \ref{Q}, we plot the evolution of the dissipation parameter during both phases of CDWI. We consider a total $50$ e-folds duration of inflation and at the pivot scale $N_P=50$, fix $Q_P=100$. It can be seen that in Phase-I $(\Upsilon\propto T^3)$, the dissipation parameter evolves slowly, while in Phase-II $(\Upsilon\propto T^3/\phi^2)$, the growth in $Q_k$ is huge. Overall, due to this amplification in the dissipation parameter, the primordial power spectrum grows tremendously on the small scales, which can lead to the formation of primordial black holes.
We will investigate PBH formation in detail in a separate paper.

	\section{Summary and Discussion}
    \label{summary}
    Warm inflation is a generalized description of inflation, where the dissipative effects in a coupled inflaton-radiation system lead to a finite temperature in the universe throughout its evolution. We discuss a composite dissipation warm inflation (CDWI) model with two terms in the dissipation coefficient: $(\Upsilon\propto T^3/M_{Pl}^2)$ contributing at the large scales and $(\Upsilon\propto T^3/\phi^2)$ dominant at the small scales. 
    As the dissipation coefficient arises from the microphysics of inflaton interactions, including the decay channels of the inflaton, coupling strengths, and multiplicities of fields, a dynamical evolution of the inflaton-radiation system can lead to a smooth transition between different functional forms of the dissipation coefficient. The exact microphysics construction of the CDWI model studied here needs to be rigorously worked out.
   Our CDWI model is characterized by two phases of inflation - Phase-I in a strong dissipative regime with a red-tilted primordial power spectrum, consistent with the recent ACT results, and Phase-II with a blue-tilted spectrum and enhanced power at the small-scales.

    In this study, we carry out a phenomenological study of CDWI model and investigate the effects of different parameters - such as the duration of Phase-I and Phase-II, the slow-roll parameter at the transition point, dissipation strength at the pivot scale, and the growth function - on the primordial power spectrum and its tilt. Our analysis shows that we require a strong dissipation regime ($Q\gg1$) and a large number of e-folds of Phase-I ($N_1>30$) to obtain the spectral index consistent with CMB observations. Further, we see that the transition parameter $x$ determines the number of e-folds of Phase-II - for $x=1/3$, we obtain $N_2=10$, and $x=1/4$ yields $N_2=15$. In this study, we consider the combined number of e-folds in the two phases to be $50$. Thus, we get two consistent CDWI models: $N_1=40$ and $x=1/3$, and $N_1=35$ and $x=1/4$. In these models, we find that the inflaton field has a sub-Planckian field excursion, implying that the swampland distance conjecture is followed. Thus, the CDWI model is also interesting from the viewpoint of embedding it in a UV complete theory.

    Further, we find that in Phase-II, the dissipation parameter grows tremendously,  which leads to a huge growth in the amplitude of primordial power.
     We find that the growth function $G_2(Q)$ is the most suitable for enhancing the power spectrum at the small-scales, sufficient for PBH formation. The formation of PBHs and the associated spectrum of induced gravitational waves also needs to be further studied.

    \section{Acknowledgements}
    RA acknowledges financial support from the MNNIT through the Seed Grant Scheme: Project No. $926/R\&C/2023$-$24$. SEJ thanks FAPERJ for the financial support.
	
\bibliographystyle{utphys}
\bibliography{main}	
\end{document}